\documentclass[a4paper,11pt]{article}
\pdfoutput=1 

\usepackage{jcappub} 

\usepackage[T1]{fontenc} 
\usepackage{diagbox}
\usepackage{float}
\usepackage{color}
\newcommand{\redn}{\color{black}}
\newcommand{\black}{\color{black}}
\newcommand{\red}{\black}

\title{\boldmath Neutrino production in blazar radio cores}

\author[a,b]{Oleg Kalashev,}
\author[c,a,1]{Polina Kivokurtseva,\note{Corresponding author.}}
\author[a,c]{and Sergey Troitsky}

\affiliation[a]{Institute for Nuclear Research of the Russian Academy of Sciences,\\ 60th October Anniversary Prospect 7a, Moscow 117312, Russia}
\affiliation[b]{Moscow Institute for Physics and Technology,\\ 9 Institutskiy per., Dolgoprudny, Moscow Region, 141701 Russia}
\affiliation[c]{Faculty of Physics, M.V. Lomonosov Moscow State University,\\ 1-2 Leninskie Gory,  Moscow 119991, Russia}

\emailAdd{kalashev@ms2.inr.ac.ru}
\emailAdd{kivokurtceva.pi19@physics.msu.ru}
\emailAdd{st@ms2.inr.ac.ru}

\abstract{Models of the origin of astrophysical neutrinos with energies from TeVs to PeVs are strongly constrained by multimessenger observations and population studies. Recent results point to statistically significant associations between these neutrinos and active galactic nuclei (AGN) selected by their radio flux observed with very-long-baseline interferometry (VLBI). This suggests that the neutrinos are produced in central parsecs of blazars, AGN with relativistic jets pointing to the observer. However, conventional AGN models tend to explain only the highest-energy part of the neutrino flux observationally associated with blazars. Here we discuss in detail how the neutrinos can be produced in the part of an AGN giving the dominant contribution to the VLBI radio flux, the radio core located close to the jet base. Physical conditions there differ both from the immediate environment of the central black hole and from the plasma blobs moving along the jet. Required neutrino fluxes, considerably smaller than those of photons, can be produced in interactions of relativistic protons, accelerated closer to the black hole, with radiation in the core.}


\arxivnumber{2212.03151}

\usepackage{amsfonts,graphics,color}
\usepackage{graphicx}
\usepackage{color}
\usepackage{hyperref}

\begin{document}
\maketitle
\flushbottom


\section{Introduction}
\label{sec:intro}
Recently, tremendous progress has been made in observation of high-energy neutrinos, with the existence of their astrophysical flux established by IceCube \cite{IceCube-2013,IceCube-HESE-2020}, ANTARES \cite{ANTARES-diffuse} and Baikal-GVD \cite{Baikal-Neutrino2022} experiments. At the same time, both the huge background of atmospheric events and poor angular resolution of neutrino telescopes make it difficult to establish their actual astrophysical sources, see e.g.\ Ref.~\cite{ST-UFN} for a recent review. 

The evidence is however growing that a significant part of high-energy neutrinos are associated with blazars, that is active galactic nuclei with relativistic jets pointing to the observer. While the first indication to the blazar origin of neutrinos came from a coincidence of a high-energy neutrino candidate event with a gamma-ray flare of a blazar \cite{IceCubeTXSgamma}, it has been shown that the population of gamma-ray loud blazars cannot provide for a sizeable contribution to the overall observed neutrino flux, see e.g.\ Refs.~\cite{PtitsynaNoGamma,Murase10percent}. A more general marker of a jet pointing to the observer is the Doppler enhancement of the radio flux, seen by very long baseline interferometry (VLBI) \red \cite{Zensus:1997ty,Blandford:jets-review}. \black Statistically significant directional associations between the ensembles of high-energy neutrino events and populations of VLBI-selected blazars have indeed been found \cite{neutradio1,neutradio2}. Other neutrino-blazar associations \cite{Resconi2020,Franckowiak2022,Buson2022} are also dominated by VLBI-bright compact sources. Newer alert events, published after the original Ref.~\cite{neutradio1}, continue to support this association \cite{neutradio2022}. Moreover, at high neutrino energies, where lower atmospheric background makes a study of this kind feasible, moments of neutrino arrival are correlated with radio flares \cite{neutradio1,Hovatta2021}.

A number of models of high-energy neutrino production in active galaxies have been developed since the early days of neutrino astronomy, see e.g.\ Refs.~\cite{Meszaros-rev,Murase-rev,AhlersHalzen,Boettcher-rev,Cerruti-rev} and \cite{ST-UFN} for reviews and extensive lists of references. However, when confronted with the observational results mentioned above, most of these models face two problems. Firstly, they predict mostly neutrino of PeV energies, while the blazar association was established \cite{neutradio2,Buson2022} for all energies from TeV to PeV. Secondly, they do not predict the observed correlation \cite{neutradio1,Hovatta2021} between the neutrino and VLBI radio fluxes. While some theoretical ideas towards explanation of the neutrino/VLBI associations have been put forward \citep{NeronovSemikozRadio,neutradio2}, a generally accepted model is missing.

In the present work, we aim to fill this gap. We recall that the dominant contribution to the VLBI radio flux comes from the compact core and adopt the mechanism sketched in Ref.~\cite{neutradio2} to physical conditions expected in this stationary feature at the base of the jet. This makes it possible to explain the neutrino/VLBI association without requiring either too long proton interaction time or too high proton power. The price we pay is the necessity to accelerate protons and to produce neutrinos in different places, in agreement with the general arguments arising in favor of multi-zone models of blazars necessary to explain all multimessenger data.

The rest of the paper is organized as follows. In Sec.~\ref{sec:prod}, we give an overview of the radio cores of blazars and of the mechanism of neutrino production we propose. We start in Sec.~\ref{sec:prod:core} with a brief review of the parts of a blazar contributing to radio observations, with a particular emphasis to the stationary core, stressing possible vagueness of the term. We then concentrate on the model of the stationary oblique shock in the millimeter radio core of a blazar and describe it in more detail in Sec.~\ref{sec:prod:shock}. Section~\ref{sec:prod:estimates} contains approximate but important estimates of the relations between proton, neutrino and photon fluxes from the core. More detailed numerical calculations of the fluxes are presented in Sec.~\ref{sec:num}. In Sec.~\ref{sec:disc}, we compare the mechanism we propose with previous studies and discuss its observational implications. We briefly conclude in Sec.~\ref{sec:concl}.

\section{Overview and estimates}
\label{sec:prod}
\subsection{Central parts of a blazar and the VLBI radio core}
\label{sec:prod:core}
In order to set up the stage for our further discussion, let us review briefly general observational results on the very central parts of blazars. Note that the sketch presented in this section is too simplistic, and many details and variations are ignored. We refer to numerous astrophysical textbooks and review papers, e.g.\ Refs.~\cite{Derm-Menon-book,Jets-book,Marscher:blazars-review,Blandford:jets-review,Hovatta:jets-review}, for more details and discussions.

In the center of a bright blazar,  there is a supermassive black hole (SMBH) with a typical mass of order $10^8$ solar masses. The radiation of a blazar is powered by accretion of matter on SMBH and is in particular related to relativistic jets fed by this accreting material. The mechanisms of jet launching are not firmly known but in any case should provide conditions for particle acceleration. The observed radiation of a blazar is strongly enhanced by Doppler boosting, because the emitting particles are moving at relativistic velocities in the direction close to the line of sight, see e.g.\ Ref.~\cite{Lister_2016}. 

The characteristic linear scale of a SMBH is given by its gravitational radius which, for the quoted value of the mass, is $R_{\rm G}\sim 10^{-5}$~pc~$\sim 0.3$~light hours. Observed intranight (in a few cases, minute-scale) variability of the gamma radiation of blazars suggests that the gamma rays probe the region very close to SMBH, probably the one where the jet is launched. The broadband spectral energy distributions of blazars have two wide bumps separated by several orders of magnitude in energy. It is generally accepted that the lower energy bump is produced by the synchrotron radiation and the higher energy one by inverse Compton scattering of relativistic electrons. However, these emissions are not necessarily produced in one and the same place (which would then coincide with the immediate vicinity of SMBH, given the gamma-ray variability).

The strongest motivation for the multi-zone blazar models comes from radio observations. At centimeter wavelengths, where the bulk of radio observations are carried out, central parts of the blazar are opaque because of synchrotron self-absorption. The observed radio emission therefore has to originate from more distant regions along the jet. The radiation observed with VLBI comes from the regions located at a few parsecs from SMBH, where the jet width is a sizeable fraction of parsec. A typical VLBI image consists of the core, a stationary source, unresolved even at the best resolution, and a few moving knots along the jet, see e.g.\ Refs.~\cite{Lister_2016,Jorstad_2017,Jorstad_2001}. For most sources, the stationary core is the brightest VLBI feature, giving the dominant contribution to the VLBI flux \cite{Jorstad_2017}. It is variable, and the increase of the core flux often precedes the separation of a new moving feature, interpreted as a plasma blob flying away from SMBH along the jet. The typical variability timescale in radio observations is months to years, consistent with sub-parsec linear scales of the physical regions being observed. Note that single-dish radio observations, with much worse angular resolution as compared to VLBI, integrate the radio flux over much larger regions of the jet, possibly incuding even kiloparsec-scale structures. The variability of the total single-dish flux, also observed at the time scales of months to years, is saturated by the emission from parsec-scale regions seen with VLBI. Therefore, less expensive to obtain single-dish light curves can serve as proxies to the VLBI core flux variability studies. 

The jet becomes transparent for radio emission of different frequencies at different distances from SMBH \cite{Blandford:jets}, and the astrometrical core position indeed changes with the wavelength, see e.g.\ \cite{Blandford:jets,CoreShift}. This leads to the interpretation of the core as the brightest visible part close to the jet base and not as a single physical object. However, at millimeter wavelengths, the jet should be transparent at the core position, and the millimeter VLBI core may be associated with a particular physical part of the jet \cite{Daly_Marscher_1988}: if even higher-frequency observations were available, the core position would not move further.

We note a potential confusion in the usage of the term ``blazar core'', which is sometimes attributed even to the entire central region of a blazar. In many cases, by the ``radio core'' one understands a frequency-dependent stationary observed feature at centimeter wavelengths, that is a part of the jet. In the present work, we use the term ``core'' for the physical part of the jet close to its base seen as a stationary spot at millimeter wavelength, that is the ``millimeter core'' in the sense of Ref.~\cite{Daly_Marscher_1988}.

\subsection{The shock model of the core, the site of neutrino production}
\label{sec:prod:shock}
According to Ref.~\cite{Daly_Marscher_1988}, the core is associated with a standing or slow shock in the region not far from the base of the jet. 
The shape of the jet's boundary is determined by two parameters: the initial bulk Lorentz factor $\Gamma_0$ and the external pressure-to-initial pressure ratio $P_e/P_0$. When the jet enters a low-pressure region, it expands too much (due to the inertia of the higher-velocity interior regions), then contracts as expansion waves are reflected off the boundary as compression waves. 
The jet recollimates, forming the conical shock.
The shock forms in the medium of electrons moving with large Lorentz factors through the region. The electrons are further accelerated at the shock and, due to the increase of the magnetic field and density by compression at the shock front, their synchrotron radiation results in the enhancement of the flux from this region, compared to that of the bulk of the jet. Although compression causes some heating, the main effect of the energy gain comes from diffusive shock acceleration, effective if the magnetic field is nearly parallel to the shock normal. Standing shocks form a very narrow angle with respect to the jet axis, $\alpha \lesssim 10^\circ$ \cite{Daly_Marscher_1988}. As a result, such shocks have the shape of narrow cones. This allows electrons to cross the shock multiple times while keeping the relativistic velocity along the jet.

While extensive VLBI observations of blazars at millimeter wavelengths, required to test this picture, are still missing, a few recent results of such observations may be explained within the frameworks of the standing-shock model of the core \cite{EHT-3C279,EHT-J1924-2914}. When a jet is viewed at an angle larger than $\sim 20^\circ$, a similar source is observed as a radio galaxy, and the beaming of the relativistic part of the jet is greatly reduced. Observations suggest that jets are layered, see e.g.\ Refs.\ \cite{Walker-2018-M87,CenA-2021-EHT,RadioAstron-M87-2023,M87-2023} for radio galaxies and Refs.\ \cite{Piner-2010-blazars,RadioAstron-2021-3c273} for blazars. What is seen in radio galaxies is dominated by a slower sheath, while the shock in the millimeter core of blazars resides in the very fast spine, see e.g.\  Refs.~\cite{Hada-et-al-2011-Nature,Marsher-Nature-comment}.

The bulk of the VLBI flux, which was found to correlate with the neutrino emission \cite{neutradio1,Hovatta2021,neutradio2}, comes from the stationary  core, not from moving knots associated with plasma blobs that are flying away. Here, we associate the neutrino production zone with this shock region, so that a direct connection between VLBI and neutrino observations is implied.

We propose a two-zone model in which protons are accelerated close to the black hole and approach the radio core having energies required for neutrino production, that is much higher than those of electrons. We do not specify the precise location of the zone where the protons are accelerated, nor a particular acceleration mechanism. The required proton energies $\sim(10^{15}\dots 10^{16})$~eV can be obtained by acceleration in the electromagnetic field close to the black hole \cite{PtitsynaNeronovGap}, by magnetic reconnection \cite{magnetic-reconnection} or by shock acceleration close to the base of the jet \cite{Bykov-accel-1205.2208}.
\begin{figure*}
\centerline{\includegraphics[trim=15 125 40 80,clip,width=\linewidth]{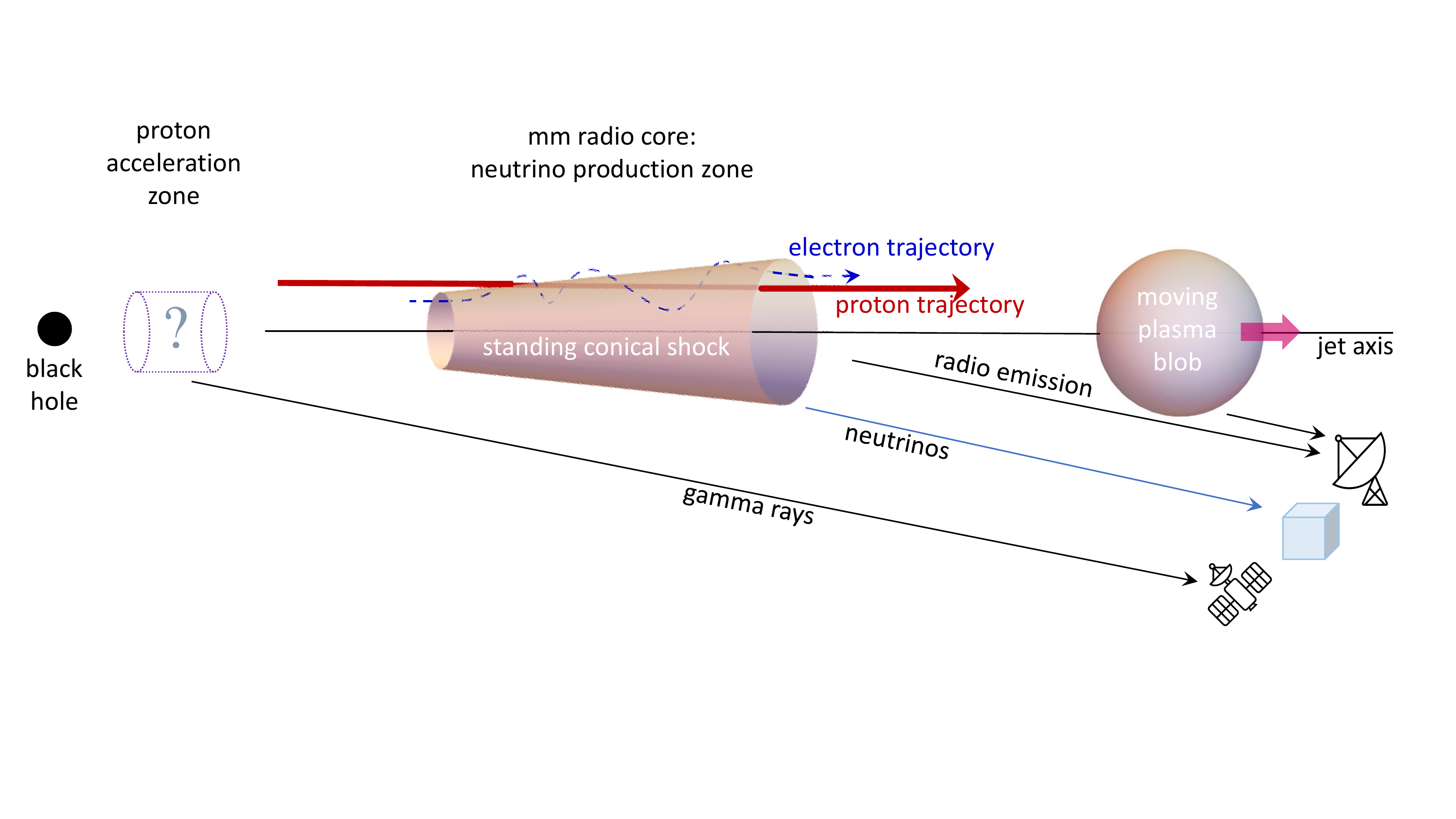}}
\caption{\label{fig:sketch} Sketch (not to scale) of central parsecs of a blazar and jet structures relevant for neutrino production and radio observations. See the text for a detailed discussion.}
\end{figure*}

\subsection{Estimates of the neutrino flux and proton power}
\label{sec:prod:estimates}

In this subsection, we present rough order-of-magnitude estimates of the neutrino flux produced by relativistic protons in the radio core of a blazar with typical parameters. Then we normalize the neutrino flux to the average neutrino flux of a blazar, estimated in Ref.~\cite{neutradio2} from observations, and obtain the required proton power.

We assume that the bulk of electromagnetic radiation of the core is associated with relativistic electrons and includes synchrotron radiation at lower energies together with inverse-Compton radiation at high energies. In the latter component, there is always the synchrotron-self-Compton (SSC) contribution, which may or may not dominate over Compton radiation on external photons. This SSC component is directly related to the synchrotron emission seen in radio with VLBI. Following Ref.~\cite{neutradio2}, we consider scattering of relativistic protons on these SSC photons and estimate the proton power required to reproduce the observed neutrino flux by means of these interactions. However, Ref.~\cite{neutradio2} assumed that the $p\gamma$ interactions take place in a relativistic spherical blob flying away along the jet. Here, we change the geometry of the emission region to that of the shock, which may be considered as a cylinder of radius $r'$ and length $l'$ for our purposes. Hereafter, primed quantities correspond to the frame in which electrons in the core, and hence their secondary SSC photons, are isotropic. 

The electrons in the core gain their velocity in the direction perpendicular to the shock, while the velocity component parallel to the shock remains mildly relativistic. Beyond this standing shock, these accelerated electrons move along the jet, that is very close to the line of sight. Therefore, one expects that the Doppler factor of the core, $\delta_c$, is smaller than that of the jet, $\delta$. There exist observational indications to this trend \cite{DopCoreJet,Lee:2016ctn,Weaver:2022qty}.

The observed flux of SSC photons $\mathcal{F}$ can be related to the concentration of the photons in the source by making use of Eqs.~(5.46), (5.47) of Ref.~\cite{Derm-Menon-book}. The only change here is the geometry of the emission region, which implies the volume $V'=\pi r'^2 l'$ instead of $(4/3)\pi r'^3$ and the light crossing time $ct'_{\rm lc} \sim \left(r'^2l'\right)^{1/3}$ instead of $r'$. We obtain
\begin{equation}
E'_\gamma n'_\gamma(E'_\gamma)=
\left( E'_\gamma n'_\gamma(E'_\gamma) \right)_b \cdot
\frac{4}{3}
\left( \frac{\delta}{\delta_c}\right)^{4}
\left( \frac{r'}{l'}\right)^{2/3},
\label{background_rescale_core}
\end{equation}
where we explicitly separate the old result \cite{neutradio2} for the blob,
\begin{equation}
    \left( E'_\gamma n'_\gamma(E'_\gamma) \right)_b =
\frac{3d_L^2\mathcal{F}}{\delta^4E'_\gamma r'^2};
\label{background_rescale}
\end{equation}
$E'_\gamma$ and $n'_\gamma$ are the energy and the energy-dependent concentration (per unit energy) of photons in the source, $\delta$ and $\delta_c$ are Doppler factors of the blob and of the core, and $d_L$ is the luminosity distance to the source. Taking benchmark values for radio blazars, $r'\sim0.2$~pc, redshift $z\sim1$, $\mathcal{F}\sim 8 \times 10^{-12}$~erg/cm$^2$/s and $\delta\sim 10$, see discussion and references in Ref.~\cite{neutradio2}, we obtain $\left( E'_\gamma n'_\gamma(E'_\gamma) \right)_b  \simeq 5.4 \times 10^3$~cm$^{-3}$ (note a misprint in Ref.~\cite{neutradio2} for this number). For the cone angle $\alpha=10^\circ$, one has $l'\approx 1.2$~pc.

 Neutrinos with observed energies $E_{\nu}$ between 1 TeV and 1 PeV contributed cumulatively to the effect found in Ref.~\cite{neutradio2}, but the result was dominated by events with $E_\nu \sim 40$~TeV.  The $p\gamma$ process in which  these neutrinos  may be produced is dominated by the $\Delta$-resonance channel. In this case, the required proton energies in the emitting-region frame are $E'_{p} \approx 20E'_{\nu} (1 + z)/\delta_c$. This also fixes the energy of the target gamma rays as $E'_\gamma \approx m_\Delta^2/E'_{p}$, that is the keV to MeV band (the flux estimate $\mathcal{F}$ used above is for these energies).  The approximate relation between the proton, $L_p$, and neutrino, $L_\nu$, luminosities reads as $L_p\sim 20L_\nu/(l'_1 \sigma n')$, where $\sigma\approx 500~\mu$b is the resonance cross section of the $p\gamma$ interaction and $l'_1$ is the proton path through the photons. The magnetic field in the core is poorly known, and while electrons are trapped in the shock, high-energy protons may or may not be trapped. We therefore assume $l'_1=l'$ for a conservative estimate: if the proton trajectory is curved, then the neutrinos would be produced more efficiently, and lower proton power would be required to match the neutrino luminosity of a blazar, $L_\nu\sim 10^{43}$~erg/s, estimated in Ref.~\cite{neutradio2} from observations. 

This average luminosity was estimated in Ref.~\cite{neutradio2} on a statistical basis, and not source-by-source, since neutrino fluxes of individual blazars are too low to be firmly detected by present instruments.

Note that Ref.~\cite{neutradio2} assumed isotropic neutrino emission in the blob frame, which is no longer valid in the present study: while photons are isotropic in the frame we are working, protons are not. However, the total $4\pi$ luminosity is a relativistic invariant, so the same value is valid for this frame as well. Since $L_p$ is not known observationally, it can be treated as a free parameter: to get more neutrinos, just add more protons. The value of the required proton luminosity, $L_p$, should not be too high for the model to be considered realistic. For the values of the parameters obtained above, we have
$$
L_p \simeq 5 \times 10^{45}~\mbox{erg/s} \cdot \delta_c^4.
$$
Assuming $\delta_c\sim\delta/5\sim 2$, we obtain the proton power of the order of the Eddington luminosity for a $M=10^9 M_\odot$ black hole, which is a realistic value, orders of magnitude lower than are assumed in some proton-jet models, e.g.\ \cite{Diltz_2015}.

\section{Numerical calculation}
\label{sec:num}

Here, we go beyond the rough estimates of Sec.~\ref{sec:prod:estimates} and perform a more detailed calculation of the neutrino flux from a benchmark blazar. We also calculate the flux of accompanying gamma rays and include important effects of electromagnetic cascades.

We perform numerical simulations using the  {\sc TransportCR} code~\cite{Kalashev:2014xna} based on transport equations. The code allows one to simulate the propagation of nuclei and nucleons through the media filled with arbitrary photon backgrounds, also tracing the secondary electron-photon cascades and neutrinos from their interactions. The background spectrum can be supplied in the table form. The code includes all relevant particle interactions for various kinematical regimes. In particular, this allows one to go beyond the $\delta$-functional approximation used in Sec.~\ref{sec:prod:estimates} for our estimates. This may be important for some astrophysical sources, see e.g.\ Ref.~\cite{Kimura}.

\begin{figure}
\center{\includegraphics[width=0.95\columnwidth]{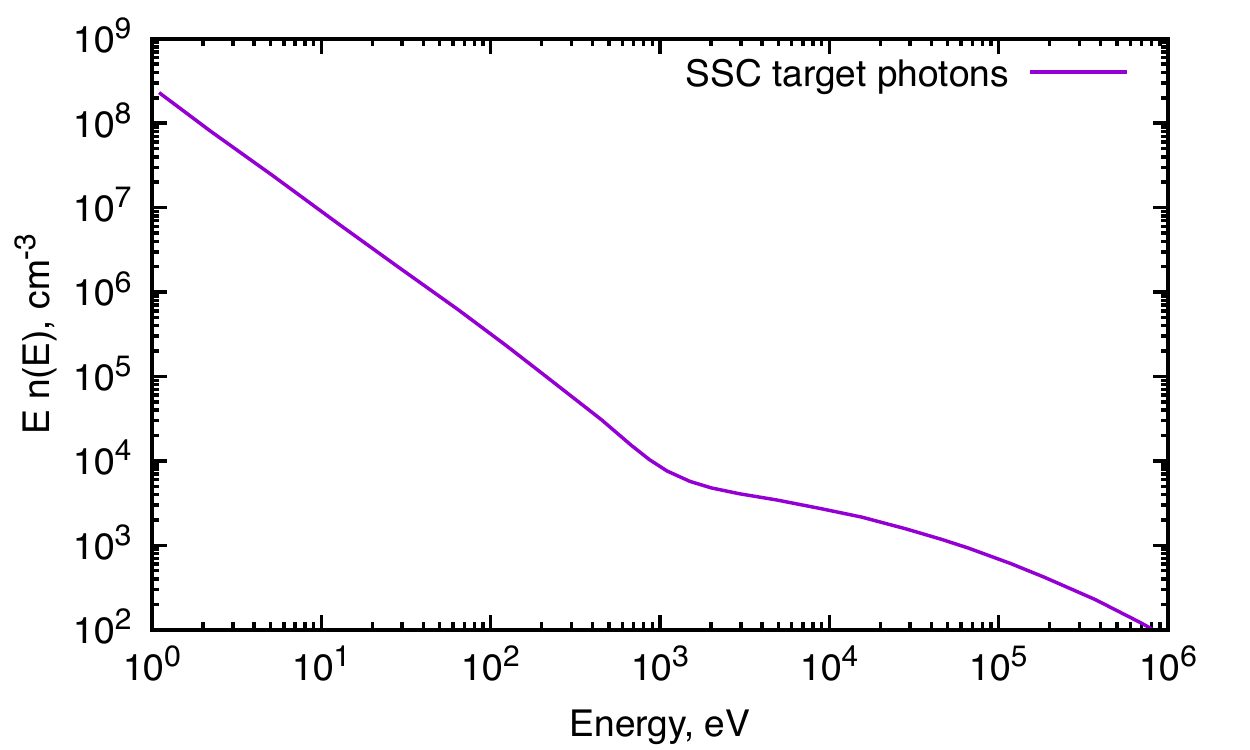}}
\caption{\redn The target-photon spectrum for 3C279 used for numerical calculations in Sec.~\ref{sec:num}.  \black}
\label{fig:target}
\end{figure}

As it was mentioned previously, SC photons act as target photons for neutrino production. For our calculations we use parameters of a well known quasar 3C~279, for which a detailed broadband spectral modelling is available. This source was associated with a high-energy neutrino event  detected by IceCube on 2015-09-26  \cite{neutradio1}. For the target photon spectrum, we use the 3C~279  synchrotron SC spectrum model from Ref.~\cite{refId0} (Fig.~7 of that work, low state), presented here in Fig.~\ref{fig:target}.
\black

We limit the target-photon background to the SSC component only, and ignore higher-energy external Compton radiation, for two reasons. Firstly, this latter component is variable at the day scale, and thus most probably comes from a more compact, compared to the radio core, source. It would be plausible to associate this second, compact zone with the site of proton acceleration, but this goes beyond the scope of our paper. The direct relevance of $\sim$~GeV gamma rays to the neutrino production in blazars is strongly constrained by the lack of the association between the Fermi-LAT blazars and neutrinos, see e.g.\ Refs.~\cite{PtitsynaNoGamma,Murase10percent}. Secondly, in any case, the effect of $\gtrsim$MeV target photons on the production of neutrinos with energies in the IceCube band is anyway small, because of the low corresponding $p\gamma$ cross section.

The simulation is performed in the core frame, so the spectrum of SSC photons of Ref.~\cite{refId0} is converted to the photon density using Eq.~(\ref{background_rescale_core}), and the resulting neutrino flux is converted back to the observer's frame using the inverse of Eq.~(\ref{background_rescale_core}). We assume the power-law injection spectrum of protons with energies  $E'_p$  up to  $10^{16}$  eV with sharp cutoff and spectral index $\alpha = 2.2$.  The choice of the maximal energy is justified by estimates in Sec.~\ref{sec:prod:estimates}, and the results are not sensitive to reasonable variations of the parameters.   For the proton propagation path we use the $l'$ estimate of Sec.~\ref{sec:prod:shock}. 
We normalize the proton flux in such a way that the resulting all-flavour neutrino luminosity of the source is $10^{43}$~erg/s as suggested by Ref.~\cite{neutradio2}. 

\begin{figure}
\center{\includegraphics[width=0.8\columnwidth]{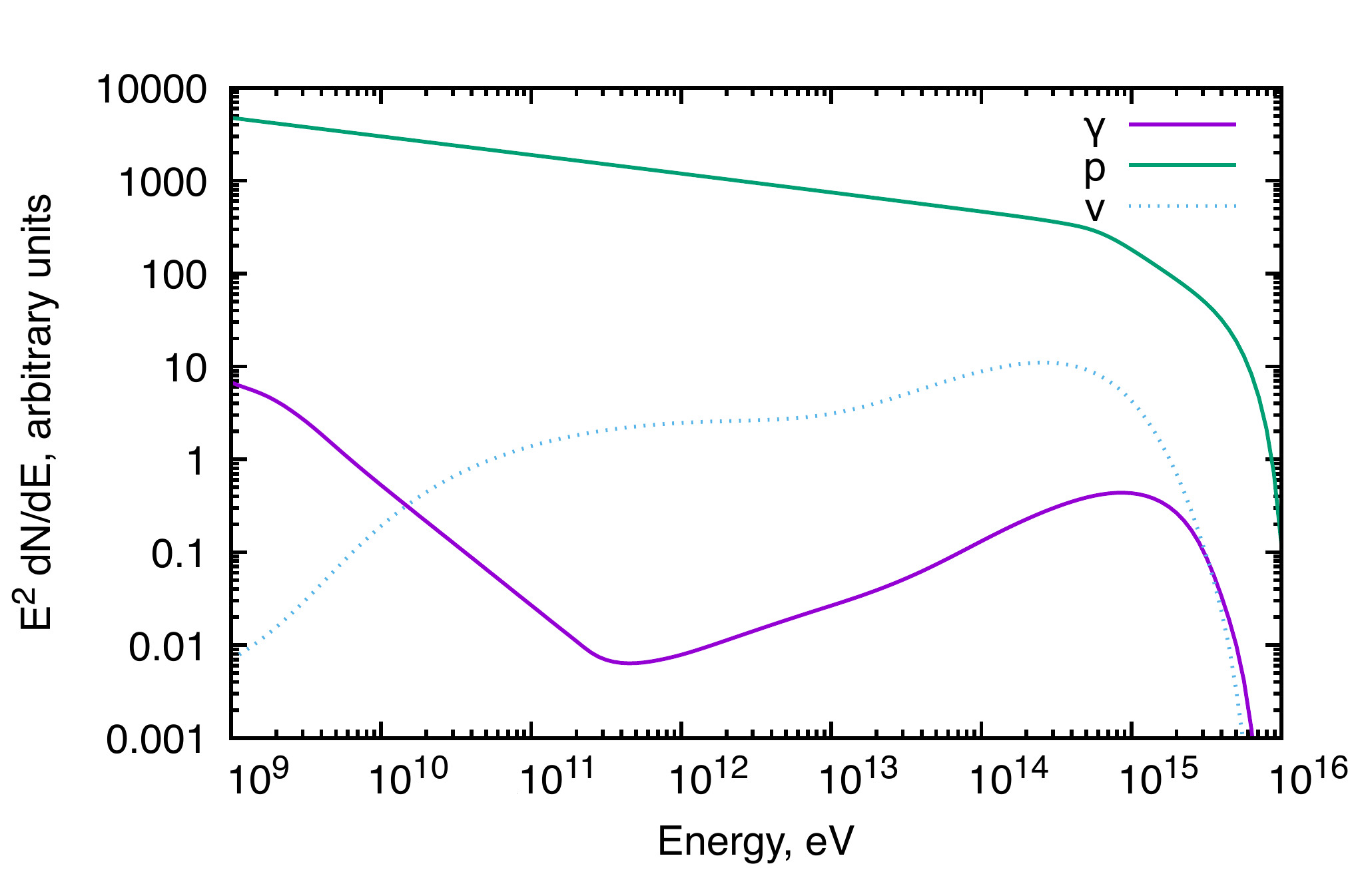}}
\caption{Prediction of the neutrino spectrum (all flavours) from the 3C279  core just out of blazar, compared to the spectra of injected protons and secondary gamma rays
, scaled to the same units. }
\label{fig:spectra}
\end{figure}

The resulting  spectrum of the produced neutrinos at the source is presented in Fig.~\ref{fig:spectra},
together with  the spectra of injected protons and of secondary  neutrino-associated  gamma rays, calculated at the exit from the source. This comparison gives an idea of the required proton power in the source, compared to the neutrino power. The calculation confirms the estimates obtained in Sec.~\ref{sec:prod:estimates}.

\begin{figure}
\center{\includegraphics[width=0.8\columnwidth]{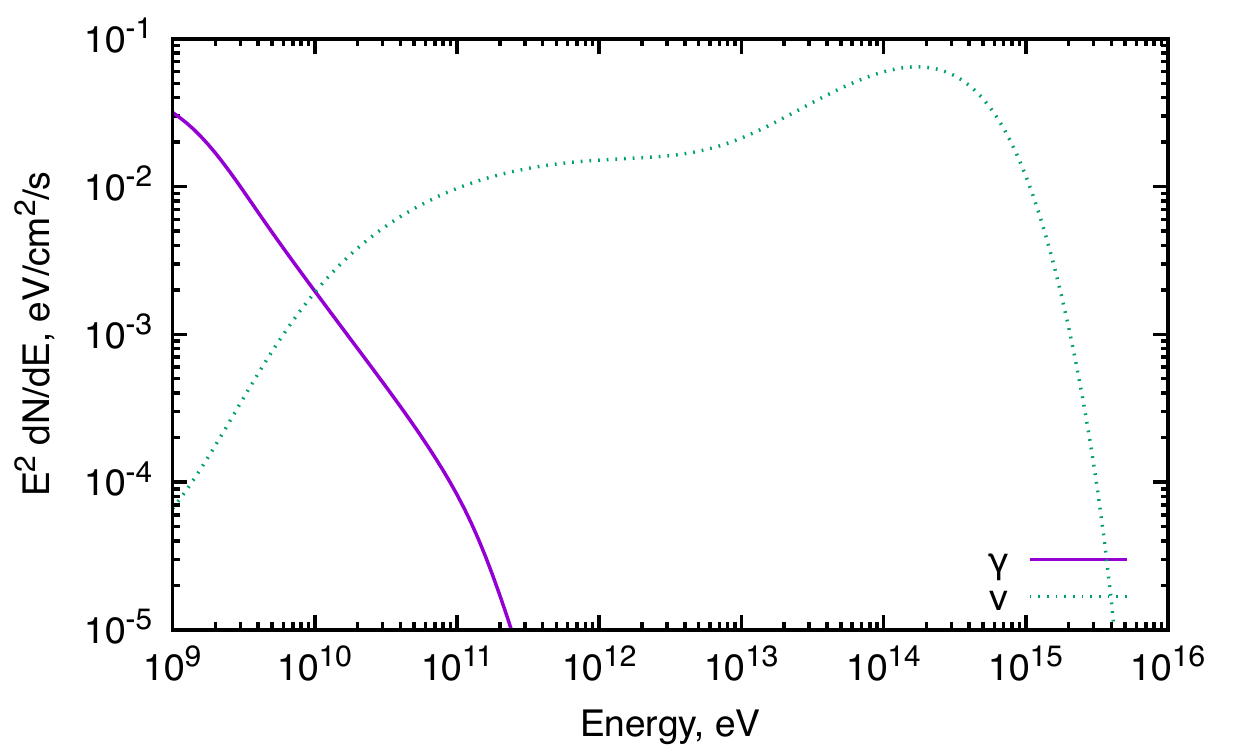}}
\caption{\redn Predictions of the observed spectra of neutrino (all flavours) and neutrino-associated gamma rays from the 3C279 core. }
\label{fig:at-Earth}
\end{figure}

Figure~\ref{fig:at-Earth}
presents the spectra of neutrinos and neutrino-associated gamma rays as they would be observed at Earth. Because of pair production on background radiation, highest-energy photons do not reach the observer, while secondary components of electromagnetic cascades do not point back to the source. The estimated GeV-band flux of the gamma rays related to the neutrino production in our model is of order \redn  $10^{-11}$~cm$^{-2}$s$^{-1}$\black, to be compared with the Fermi LAT observation of $\approx 4\times 10^{-8}$~cm$^{-2}$s$^{-1}$ flux from 3C~279. This is in line with the lack of direct correlation between neutrino and Fermi-LAT luminosities of blazars, observed e.g.\ in \cite{PtitsynaNoGamma,Murase10percent}.

\begin{figure}
\center{\includegraphics[width=0.8\columnwidth]{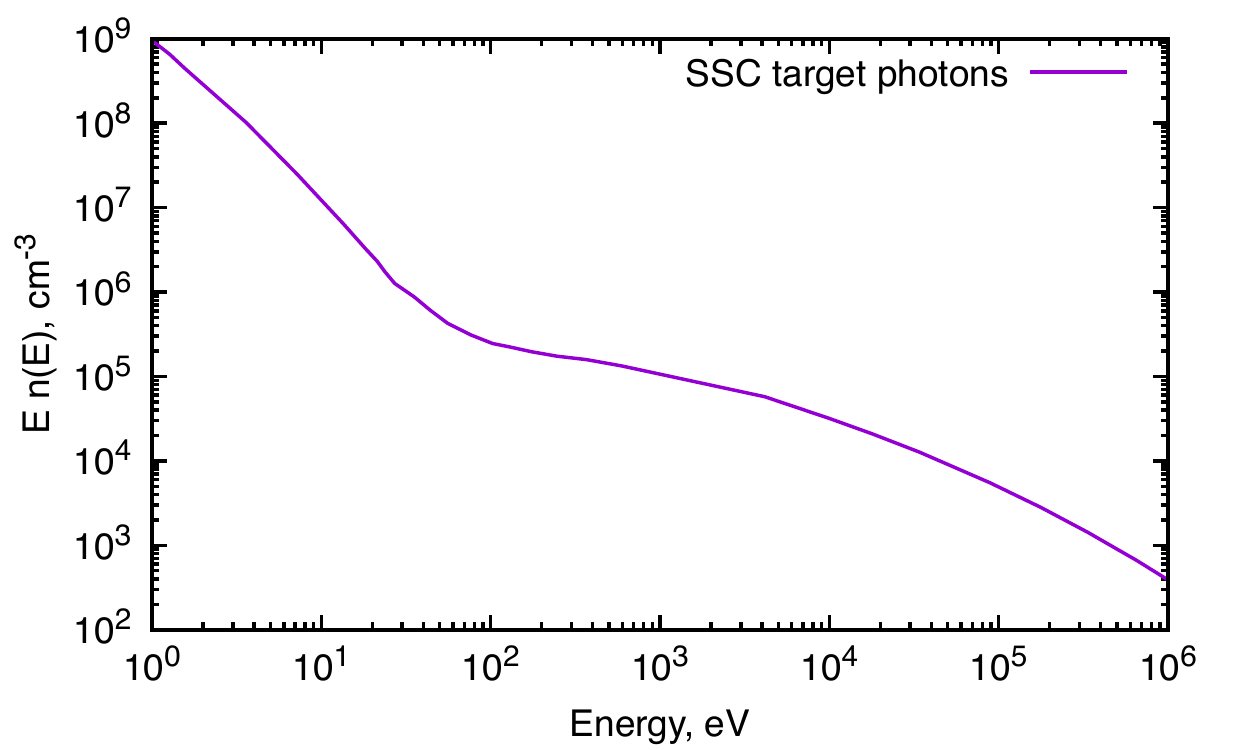}}
\caption{\redn Same as Fig.~\ref{fig:target} but for PKS~1502$+$106. \black}
\label{fig:targetPKS}
\end{figure}

\begin{figure}
\center{\includegraphics[width=0.8\columnwidth]{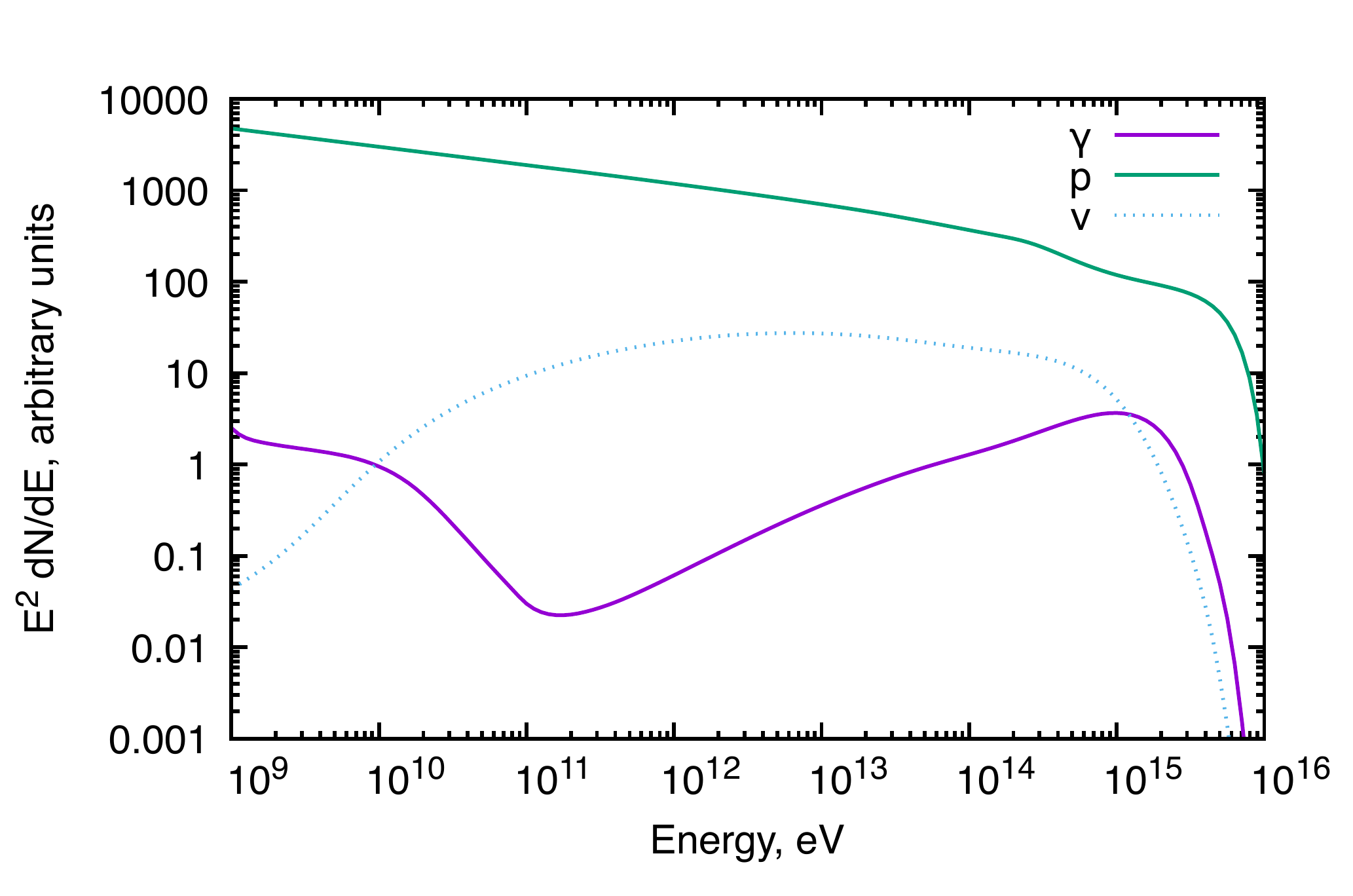}}
\caption{ Same as Fig.~\ref{fig:spectra} but for PKS~1502$+$106. }
\label{fig:spectraPKS}
\end{figure}

\begin{figure}
\center{\includegraphics[width=0.8\columnwidth]{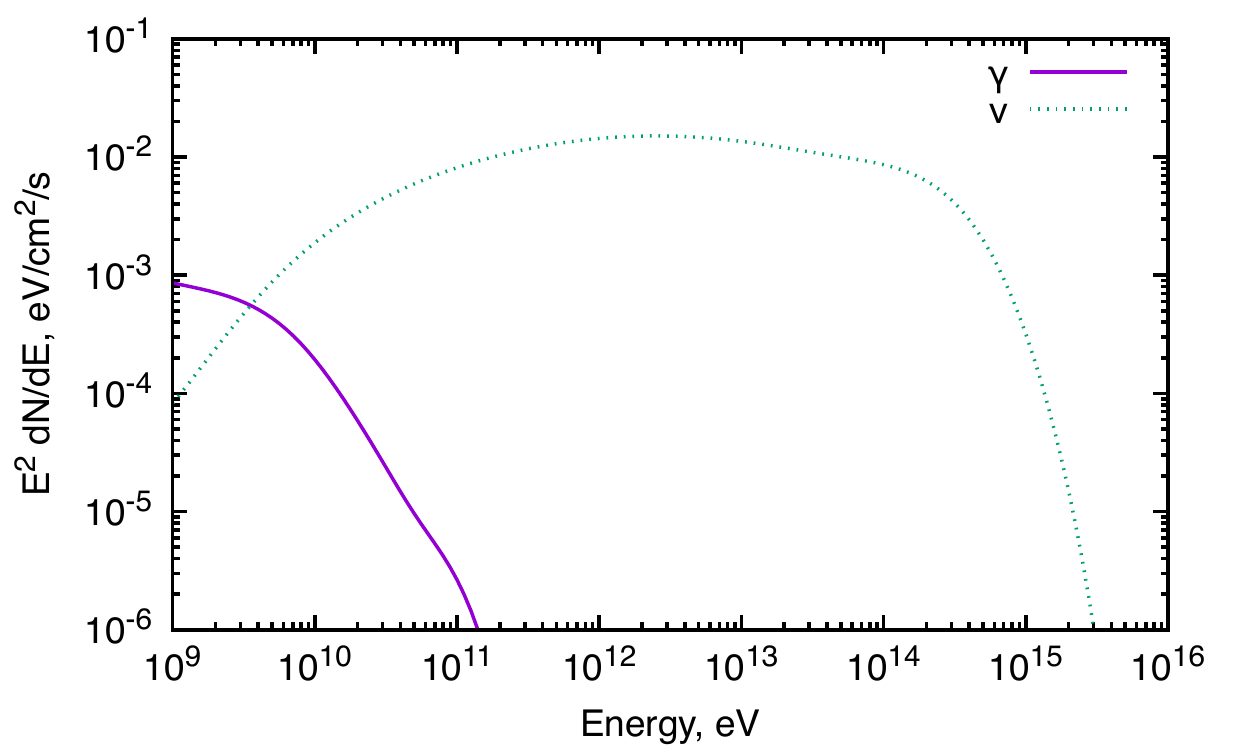}}
\caption{\redn Same as Fig.~\ref{fig:at-Earth} but for PKS~1502$+$106. \black}
\label{fig:at-EarthPKS}
\end{figure}

To illustrate how our results are affected by the diversity of blazars, we repeat our calculations for another well-studied blazar, PKS~1502$+$106, which is associated with the IceCube neutrino detected on 2019-07-30 \cite{neutradio1,1502-1,1502-2}. We use the target-photon SSC spectrum from Fig.~11 of Ref.~\cite{1502-SSC}, which we present here in Fig.~\ref{fig:targetPKS}.
Results of the calculation are presented in Figs.~\ref{fig:spectraPKS}, \ref{fig:at-EarthPKS}, 
and confirm the trends we have seen for the 3C~279 case.

\section{Discussion}
\label{sec:disc}
Various mechanisms of the neutrino production in AGN have been proposed, see e.g. Refs.~\cite{Murase-rev,Meszaros-rev,Boettcher-rev,Cerruti-rev,MuraseSteckerAGNrev} for reviews. The scenario we discuss here is very different from most of them. Indeed, the standard approach puts the neutrino production site very close to the central black hole, where high-energy gamma radiation comes from, e.g.\ Refs.~\cite{Stecker:1991vm,Stecker-1305.7404,NeronovWhich,Kalashev:AGN}. The main problem of this approach, in the context of recent observations, is the lack of keV to MeV target photons: the radiation field in this site is dominated by ultraviolet radiation from the accretion disk. These lower energies of target photons require higher proton energies for the $\Delta$ resonance to work. Consequently, the energies of produced neutrinos move towards the PeV band. This would be in contrast with results of Refs.~\cite{IceCubeTXSold, neutradio2}, where neutrinos with energies above a few TeV were associated with blazars observationally. The use of SSC photons as a target, like we adopt in the present work, resolves this tension, at the same time providing a direct link to the synchrotron radiation which dominates in the radio flux, by which the blazars were selected.

In Ref.~\cite{neutradio2}, SSC target photons were discussed, but the flying blob in the relativistic jet was considered as the site of $p\gamma$ interactions. That approach required very long path of a proton to provide for a sufficient interaction probability. This was achieved by proton trapping at the shock front and assuming the lifetime of the blob much longer than its light crossing time. Such a long lifetime, though not excluded observationally, may be challenging to achieve because of the adiabatic expansion of the blob. The radio core is very different from the blob in this aspect. First, it has $(\delta/\delta_c)^4$ times higher photon density for the same observed SSC flux. Second, its shape allows for a longer proton path in the direction of the jet. Finally, it does not expand adiabatically because particles are not confined in it and fly through the core to the jet instead. As a result, the required production rate of neutrinos is obtained naturally, without the need of very high proton luminosity. 

In the present approach, we obtain the neutrino luminosity $L_\nu$ which is orders of magnitude lower than the typical SSC photon (keV -- MeV) luminosity of the source. This relation is in a nice agreement with results of the population studies of neutrino sources based on the statistics of clustering, e.g.\ \cite{NeronovSemikozMultiplets,FinleyMultiplets}. The accompanying flux of gamma rays from the same $p\gamma$ process can be strongly suppressed because of $\gamma\gamma$ interactions in the source, thus explaining the lack of observed correlations of neutrinos with gamma-selected blazars.

Physical and observational properties of blazars are diverse. Of particular interest are the variety of bolometric luminosities and of the synchrotron peak locations in the spectral energy distribution (SED). Neutrino spectra of particular sources are therefore also different, and the numerical results presented in Sec.~\ref{sec:num} are only benchmark ones. Within our frameworks, the neutrino flux may change considerably with the change in the synchrotron peak energy, because this affects the amount of target photons in the core. One expects lower neutrino luminosities for blazars with the synchrotron peaks in the ultraviolet, because the energy required for target photons would correspond to a dip between two SED peaks in this case. Even higher-frequency objects, extreme BL Lacs, have the \textit{synchrotron} peak in the hard X-ray band and are therefore potential neutrino emitters \cite{Resconi2020}.

The neutrino flux from the core is proportional, in particular, to the amount of target SSC photons, which in turn scales with the amount of synchrotron photons. The latters manifest themselves in the radio emission from the core, detected by VLBI and used to establish the association of neutrinos with radio blazars. At the same time, the SSC photons we need have energies in the hard X-ray band, and a correlation between neutrino and hard-X-ray luminosities of blazars is also expected. It would be however less direct to establish because of contributions of different processes, e.g.\ of inverse-Compton scattering on external photons, or of unresolved but larger-scale parts of the jet, to the X-ray fluxes.

\section{Conclusions}
\label{sec:concl}
High-energy astrophysical neutrinos may be produced in ``millimeter radio cores'' of blazars, which are responsible for the dominant part of the synchrotron VLBI-detected radio flux from these AGN. To produce neutrinos with energies as low as tens of TeV, $p\gamma$ interactions require a sufficient amount of hard X-ray target photons, which are  produced by the synchrotron self-Compton mechanism in the cores. This mechanism explains naturally the association of high-energy neutrinos of different energies, from TeVs to PeVs, with radio blazars and correlations between the neutrino and radio fluxes. Certain relation between neutrinos and hard X-ray emission of the blazars is expected. Acceleration of protons should happen in a different zone closer to the central black hole, possibly the one where the gamma radiation comes from. 

The association of neutrinos with VLBI-selected blazars may however be less direct. The compact radio emission is a tracer of the Doppler enhancement of the flux, resulting from strong beaming of the relativistic jet towards the observer. Other mechanisms than $p\gamma$ \cite{Neronov:radio}, operating closer to the central black hole \cite{Riabtsev}, may be responsible for the observed correlations. Future studies of astrophysical neutrinos, together with VLBI monitoring of promising neutrino emitting blazars, are required in order to determine the mechanism and the site of neutrino production unambiguously. 

\acknowledgments
We are indebted to A.~Marscher for illuminating discussions about physical processes in the blazar radio cores. We thank the anonymous referee,  M.~Barkov, T.~Dzhatdoev, Yu.~Kovalev (Jr.), Yu.~Kovalev (Sr.), A.~Plavin and K.~Zhuravleva for helpful discussions. This work is supported by the RF Ministry of science and higher education under the contract 075-15-2020-778.
\bibliography{cores}

\end{document}